\documentclass{svjour2}                    
\smartqed  
\usepackage{graphicx}
\begin{document}

\title{Can quantum gravity be exposed in the laboratory?}
\subtitle{A tabletop experiment to reveal the quantum foam}

\titlerunning{Tabletop experiment}        

\author{Jacob D. Bekenstein         
}


\institute{Jacob D. Bekenstein \at
              Hebrew University of Jerusalem \\
              Tel.: +972-2-6584374\\
              Fax: +972-2-5611519\\
              \email{bekenste@vms.huji.ac.il}           
           \and
            \at
             }

\date{Received: date / Accepted: date}

\maketitle

\begin{abstract}
I propose an experiment that may be performed, with present low temperature and cryogenic technology, to reveal Wheeler's quantum foam.  It involves coupling an optical photon's momentum to the center of mass motion of a macroscopic transparent block with parameters such that the latter is displaced in space by approximately a Planck length.  I argue that such displacement is sensitive to quantum foam and will react back on the photon's probability of transiting the block.  This might allow determination of the precise scale at which quantum fluctuations of space-time become large, and so differentiate between the brane-world and the traditional scenarios of spacetime.

\end{abstract}

\section{Introduction}
\label{intro}
Discussion about the quantum theory of gravity has been going on for about seventy years.   But all along  it has been mostly theoretical.    Meanwhile technology in the service of physics has developed by leaps and bounds.   The recent announced discovery of the Higgs boson is just the last of a chain of experimental checks of the standard elementary particle model.  By comparison, none of quantum gravity's touted features have been put in evidence in the laboratory.   There are plenty of plans for experiments but essentially no results.  The present paper is a pedagogical reworking of the material of Ref.~\cite{JDB}, with some technical calculations suppressed.  It includes, in addition, a partial survey of previous suggestions for measuring quantum gravity effects, and a discussion of the eventuality that the here proposed experiment may give a null result.

Let us look at a couple of proposed experimental approaches to quantum gravity.   People have speculated that quantum gravity may distort the standard energy-momentum dispersion relation~\cite{Matt} $E^2=c^2 p^2+m^2 c^4$.  A parametrized distortion might be
\begin{equation}
E^2=c^2 p^2 + m^2 c^4 +\alpha\frac{E^3}{m_P\, c^2}+\beta\frac{E^4}{m_P{}^2\, c^4}+\cdots
\label{distorted}
\end{equation}
with $\alpha, \beta, \cdots$ a set of dimensionless constants. Evidently this relation clashes with Lorentz invariance, a principle that seems likely to fall victim to the introduction of quantum space-time geometry.  One consequence of the proposed distortion is that the speed of particles does not asymptote to $c$ for $E\gg mc^2$, but rather remains energy dependent.  This prediction can be tested by looking at the duration at Earth of a gamma ray burst as a function of energy.  Gamma ray bursts, comprising photons with a gamut of energies, come from very far away, and the corresponding long temporal baseline should allow the energy dependent speed to spread the burst temporally.  Since there must be an initial spread, the said measurements can only determine an upper bound on the temporal spread.  Data obtained by the Fermi $\gamma$-ray satellite show that if $\alpha$ does not vanish identically, it cannot be bigger than order unity~\cite{Fermi-LAT}.  But if $\alpha$ vanishes identically, nothing useful can be said about $\beta$.  Evidently the measurements do not settle the question posed by Eq.~(\ref{distorted}).

A second approach proposes to search for quantum black holes in the debris of collisions in the Large Hadron Collider (LHC) in CERN~\cite{Nic}.  A quantum black hole is one having a mass near Planck mass $(\hbar G/c)^{1/2}$ ($G$ is the measured Newton constant).  In reality neither LHC nor any presently imagined accelerator can access the corresponding energy of about $10^{19}\,$GeV.  What the investigators have in mind is the string-inspired scenario whereby the fermions and gauge fields which make up the matter we perceive and its interactions are confined to a four-dimensional brane (subspace) in a world with $D>4$ dimensions.  Gravity pervades the $D$-dimensional space-time~\cite{RS,DGP}.  In such 
a world the true Planck length $\ell_P^{(D)}$, the critical scale at which quantum effects become strong for gravity, is related to the nominal Planck length $(\hbar G/c^3)^{1/2}$  by~\cite{Zwiebach} 
\begin{equation}
(\ell_P^{(D)})^{D-2}=\ell_P{}^2 G^{(D)}/G
\end{equation}
 where $G^{(D)}$ is the gravitational constant in $D$ dimensions.  It is obviously possible for $\ell_P^{(D)}\gg \ell_P$.  Since $m_P^{(D)}=\hbar /(c\,\ell_P^{(D)})$, the physical Planck mass in the brane scenario can be much below $(\hbar G/c)^{1/2}$.  The LHC may thus be able to access the corresponding energy and produce black holes, but thus far no evidence of black hole formation has surfaced there.  This might mean that 
$G^{(D)}$ is not very large  compared to $G (\ell_P^{(D)})^{D-4}$, or that space-time is four dimensional after all.  The approach is obviously still not very informative.

The above are examples of large scale experiments.  Of late there have surfaced proposals for table-top experiments to probe aspects of quantum gravity.  It was remarked that a particular modification of the usual uncertainty relation in which the Planck scale figures prominently leads to modification of a number of standard quantum phenomena already at scales much larger than Planck's~\cite{Ali}.  A table-top experiment in this spirit, focusing on a Planck mass macroscopic harmonic oscillator, and employing standard techniques from quantum optics and optical interferometry, is in late stages of planning~\cite{Pikovski}.  So far there are no concrete results in this direction.

The holographic conception of spacetime has inspired an attempt to detect quantum noise in the transversal directions of a macroscopic body's path; these are related to the quantum incompatibility of the different spatial directions~\cite{Hogan}.  A proposed interferometric scheme to measure the predicted noise is under development at Fermilab, with results expected within two years. 

The dearth of experimental evidence bearing on quantum gravity has permitted many proposed theories of the subject to coexist.  Thus we have canonical quantum gravity and loop quantum gravity, both based on general relativity. We have string theory with its entourage of related scenarios such as the brane-world. We have Faddeev's ten vector theory of gravity~\cite{Faddeev}, classically equivalent to general relativity but presumably engendering a quantum gravity much different from the canonical one.  Another approach to quantum gravity is causal dynamical triangulations\cite{triang}, a numerically based scheme which views space-time as emerging from a stacking up of simplices guided by a causal rule.  Yet another tack is provided by Ho\^ rava gravity~\cite{Hor} which gives up on local Lorentz invariance to achieve the renormalizability that is lacking in canonical quantum gravity.  And there are others.

With a gamut of theoretical schemes, and almost no experimental evidence, it is well to ask if there is some accepted feature of quantum gravity which we could focus on without favoring this or that theory.  An obvious possibility is the ``quantum foam''.  Already in the 1960's Wheeler made the case that at Planck scale  quantum fluctuations of the space geometry must become strong enough to disrupt the smoothness of the space-time manifold, even to the extent of introducing multi-connected topology~\cite{Wheeler}.  On this same scale the geometry fluctuates violently in time.  Can we ``see'' this quantum foam?

The ``microscope" that could do this is very hard to come by.                 Suppose one tries to probe spacetime on Planck scale with an elementary particle. To reach a space resolution of $10^{-35}$ m, the  uncertainty principle requires the particle to have a momentum at least $10^{19}$ GeV$/c$ (the momentum uncertainty is of that order).  Now rest energies of the particles known to us,  $<10^2\,$GeV, are negligible on Planck scale, so our elementary particle will have to have an energy of about $10^{19}$ GeV.  This exceeds by eight orders the energies of the most energetic cosmic rays known, which themselves will not be imitated by accelerators for many decades. Generalizing we realize that the AchillesÕ heel of many procedures for making quantum foam graphic is in the requisite amount of localization of the probes.  The conclusion can only be one: to probe Planck scale roughness of spacetime, one has to avoid localization of the probes, either before or after the measurement.  But how can one resolve the non-smoothness of space-time without localizing the probe?

\section{An ideal quantum foam experiment}\label{sec:1}

The solution may be to arrange for translation of a macroscopic probe's center of mass (henceforth c.m.) by a controllable distance of the order Planck's length starting from an unspecified location---all this in the original rest frame of the c.m.   This relieves us of the necessity to localize the probe as part of the experiment.  How does this help us?  The idea is that an attempted shift by a distance of order Planck's length may have a different effect from shifting by a macroscopic distance.  In some theories---notably loop quantum gravity---there is no such thing as ``distances a fraction of Planck's length".   More generally one can think of the following parable.   When one drags a block of matter over a macroscopically flat surface, moving it a macroscopic distance---say, 1 cm---during a macroscopic time, one is up against dynamic friction.  If one attempts to drag it during the same time interval over only a few atomic distances, the motion is opposed by static friction---much larger than the dynamic one.  In the latter case motion is hindered by interaction between asperities in the contacting surfaces; in the former the asperities are, literally, melted away.  Reasoning by analogy, we would expect in the quantum foam case, where asperities are replaced by bumps in the metric, that the shift by a Planck length be harder to perform than a much longer shift.

As the macroscopic probe we propose an accurately rectangular block of very transparent dielectric---either amorphous or crystalline of the cubic class (which is optically isotropic). Let its dimensions be $L_1\times L_2\times L_2$, and its index of refraction $n$.  A single-photon emitter sends an optical photon accurately normally to one of the square faces.  A suitable detector records the photon after traversal of the block.  

Inside the block the photon has its momentum reduced by a factor $n$.  Why?  According to Abraham the momentum density in an electromagnetic field is~\cite{Ab,LLED} $\rho_p = (4\pi c)^{-1}|{\bf E}\times{\bf H}| =({4\pi c})^{-1}\sqrt{\epsilon/\mu}\,{{\bf E}^2}$.  The second form is specific to a plane wave in a dielectric with permittivity $\epsilon$ and permeability $\mu$; it is obtained by using the plane wave relation ${\bf H}=\sqrt{{\epsilon}/{\mu}}\,k^{-1}{\bf k}\times{\bf E}$.  For the energy density we may write $\rho_e=({8\pi})^{-1}({\epsilon {\bf E}^2}+{\mu {\bf H}^2})=(4\pi)^{-1}\epsilon {\bf E}^2$ where, again, the second form is for the plane wave.  Thus  ${\rho_e}/{\rho_p}=c\,\sqrt{\epsilon\mu}=c\,n$.  The photon is really a packet of electromagnetic fields, so the ratio of its energy to its momentum inside the block must be $c\,n$.  Outside it must be $c$.  Both outside and inside the photon energy must be $\hbar\omega$ (conservation of energy---the block is supposed immobile).  Hence, since outside the block the photon's momentum is $\hbar\omega /c$,  it is $\hbar\omega/(c\,n)$ inside the block.

What happened to the momentum difference, $\Delta p =(1-1/n)\,\hbar\omega/c$?  Momentum is conserved so momentum $\Delta p$ must have been deposited in the block, and accompanies the photon as some kind of traveling atomic excitation for the nominal duration of the  transit, $n L_1/c$.  Obviously this momentum deposit will be reclaimed by the photon upon its egress from the block.  While vested in the block the momentum $\Delta p$ endows the block's c.m. with velocity $\Delta p/M$ in the Lorentz frame in which the c.m. was originally at rest.  Thus, the c.m. is shifted by distance~\cite{Frisch,Barnett,JDB}
\begin{equation}
\Delta X_0= (L_1\hbar \omega/M c^2)(n-1)
\label{shift}
\end{equation}
before coming to rest in the said frame.  This shift is actually too small to be measured.  Rather, as we shall argue in Sec.~\ref{sec:2}, by virtue of conservation of momentum the detection of the photon beyond the block certifies that the shift specified by Eq.~(\ref{shift}) has occurred.  Note that one need not know the initial or final position of the block to know the shift.  Thus is the localization hurdle cleared.  But does  knowledge of the shift of the c.m. not clash with knowledge of the block momentum $\Delta p$?

In fact, while the commutator of c.m. position and block momentum is $\imath\hbar$, the momentum commutes with the difference of two positions; more precisely, if the block moves as a result of momentum given to it, the \textit{difference} of final and initial c.m. position operators commutes with the momentum operator~\cite{JDB}. The possession of joint knowledge of the momentum and the translation of the block is thus allowed by quantum theory.

The translation can be arranged to lie in the Planck regime; for example we take a block of high-lead glass with $L_2=5 L_1= 5\times 10^{-3}\,$m, mass density $\rho=6\times 10^3\,{\rm kg}\,{\rm m}^{-3}$ and $n=1.6$.  This has a mass $M=1.5\times 10^{-4}\,{\rm kg}$. Using a photon of wavelength $\lambda=445 {\rm nm}$ (energy $\hbar\omega= $ 2.78 eV) we get $\Delta X_0= 1.98\times 10^{-35}\, $m, quite close to the Planck length in the traditional (no large extra-dimensions) scenario.

What is the significance of such tiny translation?  Is not the c.m. just a bookkeeping device?  And how do we know that the photon momentum is vested on the full block mass rather than on a small fraction of it, which alternative would make the c.m. translation of the relevant part much larger than mentioned?  Regarding the first query one must reiterate that the c.m. position is canonically conjugate to the block momentum.    There can be no doubt about the physical significance of the momentum, so an equal physical significance accrues to the c.m. position and any shift of it.  The present account assumes that the c.m. translation has physical significance even though the c.m.---a collective coordinate---does not mark the position of any particular block constituent.  

Regarding the second query, a critic might argue that since in quantum electrodynamics a photon interacts with a single electron, the momentum given up by the photon is vested in one atom, or at most several atoms, and the full block mass is irrelevant.  In fact the photon--electron interaction leads to photon scattering, and repeated such scatterings are understood to be the reason for the dielectric's $n$ exceeding unity.   But obviously $n$, a macroscopic property, is meaningful only for a macroscopic chunk of dielectric.  Hence the full description of the phenomenon encapsulated in Eq.~\ref{shift} is no longer the local one of the critic.  The quantum electrodynamical description of the passage between the two above points of view is complex, but it is clear, as already adumbrated, that the momentum donated by the photon must accompany it as an collective (not atomic sized) excitation of the medium which propagates at speed $c/n$.  Otherwise, as it emerges from the block, the photon would not find available the required momentum to bring its own momentum back to its vacuum value.  

\begin{figure*}
  \includegraphics[width=0.75\textwidth]{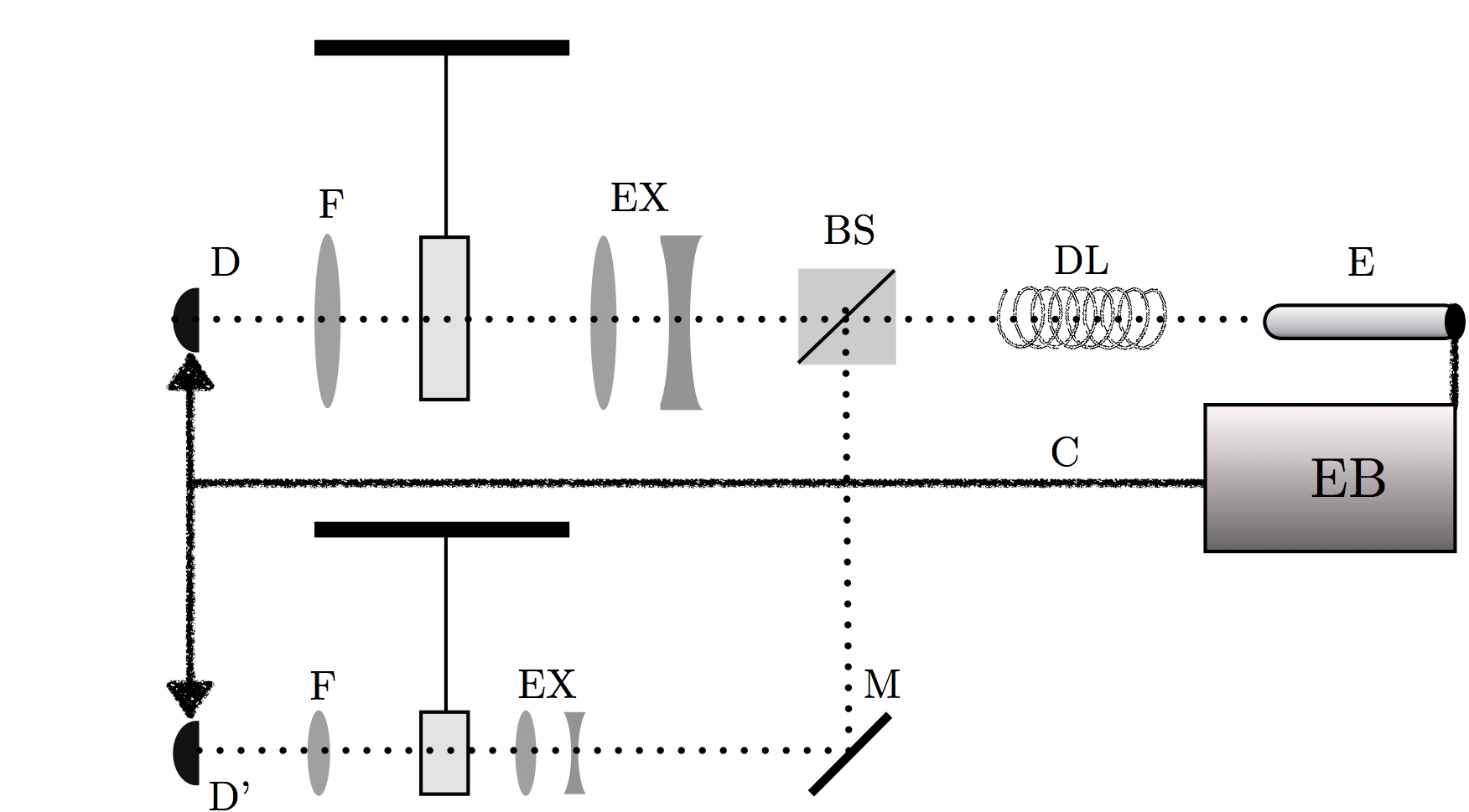}
\caption{Set up of suspended blocks showing (dotted) the alternative paths for the photon.  In the real experiment the blocks would hang side by side.  E is the single-photon emitter, D and D' are the single-photon detectors.  BS denotes the beamsplitter and M the mirror.  DL is the fiber optics delay line, and EB are the electronics that trigger D and D' through cable C.  EX are the optical elements that widen the beams while the lenses F focus them onto the detectors.  }
\label{Fig:1}       
\end{figure*}

Now if the photon beam were a thin pencil, the critic could claim that $M$ in formula~(\ref{shift}) is only the mass of a narrow central tube through the dielectric.   Thus we propose interposing,  between the single-photon generator and the block, an optical system  to spread the pulse transversally so that it ``illuminates'' most of the block (see Fig.~\ref{Fig:1}).  (A cylindrical block would evidently be more practical than a rectangular one, but for convenience sake we continue discussion with the latter.)  There is a quantum amplitude for every possible photon path through the block, but only if one actually tried to pin down the photon could one say that it has interacted with only a small part of the dielectric corresponding to a specific path.  In the absence of this meddling with it, the photon interacts with the total block mass $M$.  

\section{Planck scale motion impediment}\label{sec:2}

We return to the idea, mentioned early in Sec.~\ref{sec:1}, that motion of the c.m. over a Planck like scale is impeded.  Whereas it is rather clear intuitively that translations on scale much larger then Planck's will take place unhindered by quantum foam, this may no longer hold on scales comparable to $\ell_P$ because quantum fluctuations of the metric are large on such a scale (perturbations of order unity from the Minkowski metric).  The question is whether ballistic motion can take place when the metric has big bumps on the scale of the motion.  This is similar to asking whether free particle motion over a certain scale is possible in a potential which undulates on that scale.  The intuitive answer in both cases seems to be negative.   Let us try a different viewpoint. 

Order unity fluctuations of the metric suggest formation of virtual black holes with size similar to the fluctuation coherence length and with lifetime similar to the coherence time.  Alternatively, we can imagine moving all nonlinear terms in Einstein's equations to the right hand side, where they play the role of energy-momentum tensor of gravitation.   Because we deal with Planck scale fluctuations, on dimensional grounds the corresponding energy density must be $m_P c^2/ \ell_P{}^3$ and it has a coherence length $\ell_P$ and coherence time $\ell_P/c$.
This is sufficiently dense to form black holes of mass $m_P $ and radius $\ell_P$, really the smallest possible black holes.  One can thus envisage spacetime foam as a sea of ephemeral black holes of about Planck mass (and Planck scale radius) constantly forming and disappearing.  The block's c.m., whose translation during photon transit extends over a time very long compared to Planck's, will thus often run into such a black hole.  It seems likely that this will impede its translation.

But if the block's c.m. motion is impeded, the photon may be prevented from crossing the block because the associated transfer of momentum to the block and back to the photon in accordance with momentum conservation is not consistent with a translation smaller than $\Delta X_0$.  One cannot really argue that the momentum $\Delta p$  is transferred to the black holes instead of to the block: the gravitational vacuum must consistently have zero momentum.  To avoid contradiction the photon must do something else, e.g. get absorbed in the block or get back-reflected (the possibility of it converting into gravitons seems to have very low probability).  Since in principle one can have a very transparent dielectric, at least in the optical band, back reflection seems the more likely way out.  

I am as yet unable to estimate the probability of this anomalous back-reflection.   It could certainly turn out so small as to make the effect here described negligible.  For example, we have in the problem the Planck length and macroscopic lengths.  The probability \textit{might} come out proportional to the very small ratio between them.  In this case no effect would be detected when the elaboration of the experiment just described is carried out.  A null result would, of course, also ensue if quantum foam does not exist, or if its nature is very different from our picture of it, which, it must be stressed, is intuitive and not based on a concrete theory of quantum gravity.  Obviously these alternative reasons for a null result cannot be practically distinguished.

In this paper I am banking on a significant probability of anomalous back reflection, an eventuality which is certainly not ruled out outright.  As will become clear in Sec.~\ref{sec:3}, it should be possible to distinguish this effect from irrelevant signals.  It would certainly be an unexpected and counterintuitive effect, which in our view would reflect the existence of the quantum foam.

\section{The realistic experiment}\label{sec:3}

Our idealized set up needs to face some realities.  The block cannot be free in the lab; the best alternative is to suspend it by a very light thread.  While the restoring force originating from the tiny displacement $\Delta X_0$ is negligible, one must take into account the restoring force whose origin is in displacements occasioned by thermal noise~\cite{JDB}.  We have mentioned the use of an optical system to broaden the beam so as to encompass the full breadth of the block; there should be a second optical system to refocus the outgoing pulse onto the photon detector (see Fig.~\ref{Fig:1}).
  
In reality there is some classical back-reflection from the front and back face of the block.  One could try to avoid this complication with an antireflection coating.   But since there is always some residual reflection, and the level of the competing anomalous reflection ascribable to the quantum foam is unknown, we analyze a set-up without coating, and then describe a scheme for isolating the interesting effect from the chaff. 

According to Fresnel's formulae~\cite{LLED}, when a plane electromagnetic wave penetrates from vacuum ($n=1$) into a medium with index of refraction $n$ across a thin plane boundary, the electric field gets multiplied by $2/(n+1)$ upon transmission and by $(1-n)/(1+n)$ upon reflection.  In passage from medium to vacuum the transmission factor becomes $2 n/(n+1)$ and the reflection one becomes $(n-1)/(1+n)$.  
Thus, when the photon goes though the block with one transmission of each kind, its state picks up the factor $4n(n+1)^{-2}\exp{(\imath n\omega L_1/c)}$ which includes the phase accrued during the crossing of the block.  To this factor is associated the $\Delta X_0$ translation of the block's c.m.

But the photon can also get across by undergoing a reflection at front and back faces of the block $j$ times before finally emerging through the front face. In this case its amplitude will get multiplied by the generalization of our mentioned factor, namely,
\begin{equation}
\frac{4n}{(1+n)^2}\frac{(n-1)^{2j}}{(1+n)^{2j}}\,e^{\imath (2j+1) n\omega L_1/c}
\end{equation}
where one discerns the factor due to $2j$ reflections and the supplement to the phase from the propagation though extra distance $2jL_1/c$ inside the block.   To this event is associated a block translation by
\begin{equation}
\Delta X_j = L_1 \frac{\hbar \omega}{M c^2}(n-1+2j).
\end{equation}
Here the additional translation proportional to $2j$ results because the photon, upon each reflection at the front face of the block, conveys to the block twice it own momentum $\hbar\omega/(c n)$; it then withdraws this contribution upon reflection at the back face.  During each photon flight towards the back, which lasts time $n L_1/c$, the block has extra forward velocity $2\hbar\omega/(Mcn)$ which moves its c.m. the distance  
$2L_1\hbar\omega/(Mc^2)$.

Because the extent of translation of the block's c.m. is tied to the factor the photon's state gets multiplied by upon traversal, the c.m.'s state of the block ends up entangled with the amplitude of the photon.  Specifically, the part of the photon-block state which propagates in the direction of the photon detector (to the left in Fig.~\ref{Fig:1}) is manifestly not the product of block and photon factors:
\begin{equation}
| {\psi_\leftarrow}\rangle=\sum_{j=0}^\infty \frac{4n}{(1+n)^2}\frac{(n-1)^{2j}}{(1+n)^{2j}}\,e^{\imath(2j+1) n\omega L_1/c} \,|{\gamma_i}\rangle \otimes| {\Delta X_j}\rangle\,.
\label{ent}
\end{equation}
Here $|{\gamma_i}\rangle$ is the photon state incident on the block.
We shall not delve into the complementary back-propagating part; it is of little interest here.

We are interested in the probability that the photon gets through the block without any internal reflection; we know that this (and only this) eventuality goes together with the basic translation $\Delta X_0$.  This is $p(j=0|\leftarrow)$, the conditional probability that $j=0$ (no internal reflections) given that the photon got across.  By Bayes's theorem $p(j=0|\leftarrow)=p(j=0)/p(\leftarrow)$, where $p(\leftarrow)=\sum_{j=0}^\infty p(j)$ (this last sum is less than unity since there is some probability for the photon to be back-reflected toward the photon generator).  

From Eq.~(\ref{ent}) we see that
\begin{equation}
p(j)=\frac{16n^2}{(1+n)^4}\frac{(n-1)^{4j}}{(1+n)^{4j}}\qquad j=0,1,2,\cdots
\end{equation}
It immediately follows that $p(\leftarrow) = 2n/(n^2+1)$ so that
\begin{equation}
p(j=0|\leftarrow)=\frac{8 n (n^2+1)}{(n+1)^4}\,.
\end{equation}
For our example $n=1.6$ and $p(j=0|\leftarrow)= 0.997$.  Thus \textit{if} the photon is detected (assuming high detector quantum efficiency), the probability that the block was translated by just $\Delta X_0$  is very close to unity. 

The stress on ``if'' in the preceding statement  stems from the fact the probability of the photon transiting the block is not that high: $p(\leftarrow)=0.899$ for $n=1.6$.  Accordingly, the na\" ive probability of back-reflection of the photon is about 10\% in our example.  The essence of our claim in Sec.~\ref{sec:2} is that when the experimental parameters make $\Delta X_0$ close to $\ell_P$, this back-reflection probability is slightly enhanced because sometimes the block does not translate forward in response to the passage of the photon.  In such eventuality the photon is prevented from passing through because, by momentum conservation,  this would have entailed full translation of the block.

Probably, the anomalous increase in back-reflectivity is tiny; it may not be practical to measure it directly.  We therefore suggest the following strategy.  Along with the original suspended block and its associated detector D, suspend a second  block of identical makeup and like thickness, but considerably lower mass, followed by a separate photon detector D'.  The beam  from the photon emitter is to be directed through a beam-splitter and mirror assembly as shown in Fig.~1.  If the photon is incident on the first block, it will attempt to cause a translation of it by about $\ell_P$, and if successful it will be detected by D.  If instead it goes through the second, lighter, block, and is detected by D', it will have shifted that block a distance considerable longer than $\ell_P$. The impediment we alluded to should be all but absent under this new circumstance, so the tendency to back-reflect the photon would be felt only for the first alternative.  Since the construction of the two blocks is such as to endow them with the same Fresnel (classical) back-reflectivity, as the single-photon experiment is repeated many times we expect the detector D to detect fewer photons than D'.  This assumes that the two detectors have exactly equal quantum efficiency, and that the beam-splitter$/$mirror assembly is unbiased between the two outputs.

The last caveat is a key one.  Relative calibration of the two block setups is thus required to exclude the possibility that slight inherent asymmetry in the properties of the two blocks, or of the beam-splitter$/$mirror assembly, masquerades as a physical effect.  This could be done by preceding the envisaged series of single-photon experiments with  a separate  experiment in which a macroscopic laser pulse is sent down the same paths, and the relative intensities reaching D and D' are accurately measured. A suitable correction for any discovered asymmetry can be applied to the primary experiment.  A further precaution is relevant if one wants to analyze the one-photon events one-by-one as opposed to just looking at means and variances of multi-event series.  To prevent confusion between a click in one of the detectors (or lack of it) and a single photon emission unrelated to it, one would allow the emission to trigger the detectors through a cable issuing from electronics which detects the emission.  To give the electronics time to precede the photon, one may store the latter in a delay line, as shown in Fig.~\ref{Fig:1}. 
 
\section{Noise and its suppression}\label{sec:4}

What sources of noise might spoil the experiment?  One might think of cosmic rays, dark matter particles and solar neutrinos hitting the blocks as possible spoilers.  All these have been analyzed in the primary publication~\cite{JDB} and found to be innocuous.  While the impulses any of these corpuscles would impart to either of the blocks during or somewhat preceding a photon transit would destroy the evidence for the sought phenomenon, the chance for the passage of one of the single photons to coincide with the arrival of one of the celestial particles turns out to be very small.  Accordingly, most of the single-photon events are unspoiled.

Critical to the correct analysis of these and other noise sources is the expectation that the general character of the quantum foam remains unchanged under small Lorentz transformations such as one would expect from thermal agitation of the blocks.   While one cannot vouch for Lorentz invariance of quantum gravity in general, and of the quantum gravity vacuum in particular, if there be a preferred frame for the last, it is likely to be offset from the typical Earth-bound frame by a few times $10^5\,{\rm m\ s}^{-1}$ (the speed of the Galaxy with respect to nearby cosmological large structures).  By contrast the span of thermal velocity of a block relative to the lab is  a mere $10^{-8}\, {\rm m\ s}^{-1}$.  Thus when comparing the motion imparted to a block by the photon's passage with motion due to a noise source, it is sufficient to consider the last in the reference frame in which the block's c.m. was at rest at photon's ingress.

The most insidious noise for the present experiment comes from thermal photons in the cavity enclosing the blocks (an enclosure necessary in any case to keep out environmental optical photons).  This noise cannot be suppressed by operating in a ultrahigh vacuum since the thermal photon density is pressure independent.  At liquid helium temperature T=4\,K, the first block will be hit by no fewer than 140 thermal photons during the passage of an optical single-photon.  The thermal photons are in the microwave regime; they are scattered by the block and convey to it a momentum whose mean scales up with the square root of the number of scattered photons.  This mean momentum transfer during transit is about 1\% of the momentum temporarily deposited in the block by the optical photon~\cite{JDB}.  One may thus be tempted to consider this a tolerable level of noise.  However, the momentum due to the thermal photons is not withdrawn when the optical photon exits; after transit the block  is  moving permanently with respect to that reference frame in which its c.m. was at rest upon optical photon ingress, making determination of the translation ascribable to the optical photon's passage ambiguous.

The problem is solved by performing the experiment at 0.5\,K or below.  Because thermal photon density scales like $T^3$, and Rayleigh's scattering cross-section like the fourth power of frequency (and so like $T^4$), the thermal photon scattering rate is reduced by seven orders of magnitude, and the probability that even a single scattering is registered during optical photon transit is reduced to $10^{-5}$.  Thermal photons become irrelevant~\cite{JDB}.

Second in importance is thermal jitter of the block's c.m. due to impacts on the block's surface by residual gas atoms within the mentioned cavity.   Using off-the-shelf equipment it is possible today to reach pressures as low as $10^{-11}\,{\rm Pa}$.  Then at  4\,K the probability that the first block in our set-up is hit by just a single atom during optical photon transit is a mere $10^{-2}$.  Since thermal photon noise will necessitate even lower temperature, it is clear that most optical single-photon events will be unaffected by thermal noise~\cite{JDB}.  We refer the reader to the primary publication~\cite{JDB} for other
details of the noise analysis.

\begin{acknowledgements}
I thank the participants of the ``Horizons of Quantum Physics'' workshop in Taipei, in particular Wei-Tou Ni, Lajos Djosi and Thomas Jennewein, for useful criticism, and Al Schwartz for advice.  The present account was prepared with support from the I-CORE Program of the Planning and Budgeting Committee and the Israel Science Foundation (grant No. 1937/12), as well as from the Israel Science Foundation personal grant No. 24/12.
\end{acknowledgements}



\end{document}